\documentclass[pre,showpacs,byrevtex,amsmath,amssymb,nofootinbib,preprint]{revtex4}
\usepackage{graphicx}% Include figure files
\usepackage{dcolumn}% Align table columns on decimal point
\usepackage{bm}% bold math
\begin{document}
%\draft
\preprint{PREPRINT}

\title[Short Title]{Behavior of block-polyampholytes near a charged surface}
% Force line breaks with \\

%%%%%%%%%%%%%%% AUTHORS %%%%%%%%%%%%%%%%%%%%%%%%%%%
%1st Author
\author{Ren\'e Messina}
%\thanks{Also at Physics Department, XYZ University.}%Lines break automatically or can be forced with \\
\email{messina@thphy.uni-duesseldorf.de}% REVTEX 4
\affiliation
{Institut f\"ur Theoretische Physik II,
Heinrich-Heine-Universit\"at D\"usseldorf,
Universit\"atsstrasse 1,
D-40225 D\"usseldorf,
Germany}

\date{\today}% It is always \today, today, but you may specify any date with

\begin{abstract}
The behavior of  polyampholytes near a charged planar surface is studied by 
means of Monte Carlo simulations. The investigated polyampholytes are overall 
electrically neutral and made up of oppositely charged units (called blocks) 
that are highly charged and of the the same length.
The influence of block length and substrate's surface-charge-density on the adsorption
behavior is addressed.
A detailed structural study, including local monomer concentration, monomer mean height,
transversal chain size, interface-bond orientation correlation, is provided. 
It is demonstrated that adsorption is favored for long enough blocks and/or high enough 
Coulomb interface-ion couplings. 
By explicitly measuring the chain size in the bulk, 
it is shown that the charged interface induces either a swelling or a
shrinkage of the transversal dimension of the chain depending, in a non trivial manner, 
on the block length.  
\end{abstract}
\maketitle

%%%%%%%%%%%%%%%%%%%%%%%%%%%%%%%%%%%%%
\section{Introduction}
%%%%%%%%%%%%%%%%%%%%%%%%%%%%%%%%%%%%%

Polyelectrolytes (PEs, i.e., charged polymers) containing oppositely charged functional 
groups (blocks) are called polyampholytes (PAs). 
A well known example is provided by proteins (in particular gelatin).
This class of material can be advantageously used as adsorbate
and like that offer many applications in the every day life, such as paper production,
dewatering \cite{Watanabe_Langmuir_1999}, 
photographic films \cite{Dreja_JColIntSc_1997}, etc.

In the last decade, many experimentalist researchers have investigated the adsorption behavior
of PAs at charged substrates. For example, Neyret et al. \cite{Neyret_JColIntSc_1995}
found that PAs with a negative net charge can adsorb onto negatively charged latex particles
in accordance with the theoretical prediction of Joanny \cite{Joanny_JPhysII_1994}.
More recently, using silicon substrates Mahltig et al. 
\cite{Mahltig_JColIntSc_2001,Mahltig_JPolSciB_2001}
studied the influence of the pH on the PA adsorption properties, 
where the net charge of the PAs can be tuned in a well controlled manner.

From a theoretical viewpoint, the behavior of PAs near a
charged surface has been studied by various authors  
\cite{Joanny_JPhysII_1994,Dobrynin_Macromol_1997,Zhulina_EPJE_2001,Dobrynin_PRE_2001}.
Joanny \cite{Joanny_JPhysII_1994} showed that PAs with non-zero net charge 
can adsorb on a surface that carries the same sign of electric charge for two distinct regimes:
high and low screening strengths.
Later, Dobrynin, Rubinstein and Joanny \cite{Dobrynin_Macromol_1997}, using scaling law
arguments, sketched  typical adsorption regimes for salt-free environments.
%``pole'', ``fence'', and ``pancake'' regimes. 
%It turns out that our simulations
%will qualitatively reproduce the first and the third  of these regimes.
%
Based on a similar approach to that employed in Ref. \cite{Dobrynin_Macromol_1997}, 
scaling theories were developed to elucidate the adsorption process of a single PA chain
\cite{Zhulina_EPJE_2001} and many PA chains on a charged sphere \cite{Dobrynin_PRE_2001}.
Shusharina and Linse \cite{Shusharina_EPJE_2001} examined {\it grafted} diblock PAs
at {\it uncharged} surfaces by a mean-field lattice theory and characterized the brush structure.

As far as computer simulations are concerned, only a few studies 
\cite{McNamara_JCP_2002,Akinchina_Langmuir_2004}
were devoted to the problem of PA adsorption.
Akinchina et al. \cite{Akinchina_Langmuir_2004} performed Monte Carlo (MC) simulations 
on diblock PAs at uncharged spherical particles. 
The brush structure as a function of the charge ratio 
of the two blocks was examined \cite{Akinchina_Langmuir_2004}.
Closely related to our problem, MC simulations were carried out by McNamara et al. 
\cite{McNamara_JCP_2002} to characterize the molecular pattern recognition setting in
upon adsorption of {\it sequenced} PEs (i.e., charge-ordered PAs or non-random PAs)
to planar patterned surfaces. Using screened Coulomb interactions, they concluded that
chain-entropy prevents efficient pattern recognition \cite{McNamara_JCP_2002}. 
However, a detailed structural analysis and especially the monomer concentration profiles 
were not provided in Ref. \cite{McNamara_JCP_2002}
so that a clear description of the adsorption process of non-random PAs onto charged surfaces 
is still missing.
The fact that only patterned surfaces (checkerboards and stripes) were considered in 
Ref. \cite{McNamara_JCP_2002} makes also the process of adsorption more specific and complicated.

The goal of this paper is to provide the mechanisms governing the adsorption of
sequenced PEs (with a {\it zero} net charge) containing highly charged blocks onto charged planar surfaces.
One can already anticipate and say that the overall chain adsorption properties are 
influenced by the following driving forces:
\begin{enumerate}
\item[(i)] Chain-entropy tends to keep the chain in the solution.
\item[(ii)] Electrostatic block-block correlations tend to collapse the chain into a globular form.
\item[(iii)] A strong electrostatic interface-ion coupling favors chain adsorption.
\end{enumerate}
We limit our study to the most simple situation where a single PA chain, whose blocks
are equisized and carry the same magnitude of electric charge, interacts with a charged planar interface.
That way, it should be easier to identify the relevant mechanics setting in PA chain adsorption.
Computer simulations are especially useful in this regime of charge since most analytical theories 
\cite{Joanny_JPhysII_1994,Dobrynin_Macromol_1997,Zhulina_EPJE_2001,Dobrynin_PRE_2001}.
are only suitable for {\it weakly} charged blocks.
Our article is organized as follows: The simulation model is detailed in 
Sec. \ref{ Sec.simu_method}. Results are presented in Sec. \ref{Sec.Results},
and concluding remarks are provided in Sec. \ref{Sec.summary}.  

%%%%%%%%%%%%%%%%%%%%%%%%%%%%%%%%%%%%%
\section{Model and Parameters
\label{ Sec.simu_method}}
%%%%%%%%%%%%%%%%%%%%%%%%%%%%%%%%%%%%%

The model system under consideration is similar to that recently
investigated for the adsorption of fully positively charged PEs
\cite{Messina_PRE_2004,Messina_JCP_2006}.
Within the framework of the primitive model we consider a {\it regular}
polyampholyte near a charged hard wall with an implicit solvent 
(located at $z>0$) of relative permittivity $\epsilon_{r}\approx 78$.
To avoid the appearance of image forces \cite{Messina_JCP_2006} we suppose that 
the substrate (located at $z<0$) below the interface (at $z=0$) 
has the same dielectric constant.

The \textit{negative} bare surface-charge density of the substrate's interface is $-\sigma_0 e$,
where $e=1.60219 \times 10^{-19}{\rm C}$ is the (positive) elementary charge and $\sigma_0$ is the number 
of charges $(-e)$ per unit area. 
The latter is always electrically compensated by its accompanying 
monovalent counterions of charge $Z_+e$ 
(i.e., cations with $Z_{+}=+1$) and diameter $a$.
The block PE corresponding to a perfectly regular polyampholyte is made up of 
$2N_b$ alternating positively/negatively charged blocks:
Each block contains $M$ monovalent charged monomers of the same sign of charge
with a (typical) diameter $a$.  
Hence by construction the PA chain, containing altogether $N_m=2N_bM$ charged monomers,
is globally electroneutral. 
The corresponding counterions [i. e., monovalent anions ($Z_-=-1$) 
and cations] are also taken explicitly into account. 
Thereby, all the constitutive microions are monovalent and monosized with diameter $a$.
All those particles are immersed in a rectangular $L \times L \times  \tau$ box.
Periodic boundary conditions are applied in the $(x,y)$ directions, 
whereas hard walls are present at $z=0$ (location of the charged interface) 
and $z=\tau$ (location of an \textit{uncharged} wall).
It is to say that we work in the framework of the primitive cell model.

The total energy of interaction of the system can be written as

\begin{eqnarray}
\label{eq.U_tot}
U_{tot} & = &  
\sum_i  \left[ U_{hs}^{(intf)}(z_i) + U_{coul}^{(intf)}(z_i) \right] 
\\ \nonumber
&& + \sum _{i,i<j} \left[ U_{hs}(r_{ij}) + U_{coul}(r_{ij})
+ U_{FENE}(r_{ij}) + U_{LJ}(r_{ij}) \right],
\end{eqnarray}
where the first (single) sum stems from the interaction between a microion $i$ 
and the charged interface, 
and the second (double) sum stems from the pair interaction between microions 
$i$ and $j$ with $r_{ij}=|{\bf r}_i - {\bf r}_j|$.
All these contributions to $U_{tot}$ in Eq. (\ref{eq.U_tot})
are described in detail below.

Excluded volume interactions are modeled via a hardcore potential 
\cite{note_HS} defined as follows
%
%%%%%%%%%%%%%%%%%%%%%%%%%%%%%%%%%%%%%%%%%
\begin{equation}
\label{eq.U_hs}
U_{hs}^{(mic)}(r_{ij})=\left\{
\begin{array}{ll}
0,
& \mathrm{for}~r_{ij} \geq a \\
\infty,
& \mathrm{for}~r_{ij} < a 
\end{array}
\right.
\end{equation}
%%%%%%%%%%%%%%%%%%%%%%%%%%%%%%%%%%%%%%%%%
%
for the microion-microion one, and
%
%%%%%%%%%%%%%%%%%%%%%%%%%%%%%%%%%%%%%%%%%
\begin{equation}
\label{eq.U_hs_plate}
U_{hs}^{(intf)}(z_i)=\left\{
\begin{array}{ll}
0,
& \mathrm{for} \quad a/2 \leq z_i \leq  \tau - a/2 \\
\infty,
& \mathrm{otherwise}
\end{array}
\right.
\end{equation}
%%%%%%%%%%%%%%%%%%%%%%%%%%%%%%%%%%%%%%%%%
%
for the interface-microion one.

The pair electrostatic interaction between two microions $i$ and $j$ reads
%
%%%%%%%%%%%%%%%%%%%%%%%%%%%%%%%%%
\begin{equation}
\label{eq.U_coul} 
\beta U_{coul}^{(mic)}(r_{ij}) =
Z_i Z_j \frac{l_B}{r_{ij}} ,
\end{equation}
%%%%%%%%%%%%%%%%%%%%%%%%%%%%%%%%%
%
where $l_{B}=\beta e^{2}/(4\pi \epsilon _{0}\epsilon _{r})$ is the Bjerrum
length corresponding to the distance at which two charges $e$
interact with $1/\beta=k_B T$. 
The electrostatic potential of interaction between a microion $i$ and the
(uniformly) charged interface reads
%
%%%%%%%%%%%%%%%%%%%%%%%%%%%%%%%%%
\begin{equation}
\label{eq.U_coul_plate} 
\beta U_{coul}^{(intf)}(z_i) =
Z_i \frac{z_i}{\lambda},
\end{equation}
%%%%%%%%%%%%%%%%%%%%%%%%%%%%%%%%%
%
where $\lambda=\frac{1}{2\pi l_B \sigma_0}$ is the so-called Gouy-Chapmann length. 
The latter has two physical meanings:
(i) It is the desorption length for an isolated 
ion of charge $e$ (initially at contact with an oppositely charged wall) 
yielding an energy penalty of $k_B T$; (ii)  It also 
corresponds (in the Poisson-Boltzmann theory) 
to the distance from the wall at which surface charges are half-compensated by their counterions.
In order to characterize the electrostatic interface-ion coupling
we consider the dimensionless ``Moreira-Netz'' parameter $\Xi=\frac{l_B}{\lambda}$ 
\cite{Moreira_EPL_2000}.
A modified Lekner sum was utilized to compute 
the electrostatic interactions with periodicity in two directions.
To link our simulation parameters to experimental systems 
we choose $a =4.25$ \AA\ leading to the Bjerrum length of water
$l_{B}=1.68a =7.14$ \AA\  at $T=298$K. 

The  PA chain connectivity is modeled by a standard
finite extendible nonlinear elastic (FENE) potential for good solvent,
which reads
%
%%%%%%%%%%%%%%%%%%%%%%%%%%%%%%
\begin{equation}
\label{eq.U_fene}
U_{FENE}(r)=
\left\{ \begin{array}{ll}
\displaystyle -\frac{1}{2}\kappa R^{2}_{0}\ln \left[ 1-\frac{r^{2}}{R_{0}^{2}}\right] ,
& \textrm{for} \quad r < R_0 \\ \\
\displaystyle \infty ,
& \textrm{for} \quad r \geq R_0 \\
\end{array}
\right.
\end{equation}
%%%%%%%%%%%%%%%%%%%%%%%%%%%%%%
%
with $\kappa = 27k_{B}T/ a^2$ and $R_{0}=1.5 a$.
The excluded volume interaction between chain monomers is taken into
account via a shifted and truncated Lennard-Jones (LJ) potential given
by
%
%%%%%%%%%%%%%%%%%%%%%%%%%%%%%%%%%%%%
\begin{equation}
\label{eq.LJ}
U_{LJ}(r)=
\left\{ \begin{array}{ll}
\displaystyle
4\epsilon \left[ \left(\frac{a}{r}\right)^{12}
-\left( \frac{a}{r}\right) ^{6}\right] +\epsilon,
& \textrm{for} \quad r \leq 2^{1/6} a \\ \\
0,
& \textrm{for} \quad  r > 2^{1/6} a
\end{array}
\right.
\end{equation}
%%%%%%%%%%%%%%%%%%%%%%%%%%%%%%%%%%%%
%
where $\epsilon=k_BT$.
These parameter values lead to an equilibrium bond length $ l=0.98a$.

%
%%%%%%%%%%%%%%%%%%%%%%%%%%%%%%%%%%%%%%%%%%%%%%%%%%%%%%%%
% TABLE 1
\begin{table}
\caption{
List of key parameters with some fixed values.
}
\label{tab.simu-param}
\begin{ruledtabular}
\begin{tabular}{lc}
 Parameters&
\\
\hline
 $T=298K$&
 room temperature\\
 $\sigma_0 L^2$&
 Numbers of charge ($-e$) on the substrate\\
 $Z_{\pm}=\pm 1$&
 microion valence\\
 $a =4.25$ \AA\ &
 microion diameter\\
 $l_{B}=1.68a =7.14$ \AA\ &
 Bjerrum length\\
 $L=35 a $&
 $(x,y)$-box length\\
 $\tau=105 a $&
 $z$-box length\\
 $M$&
 number of monomer(s)  per positive/negative block\\
 $N_m=64$&
 total number of monomers\\
\end{tabular}
\end{ruledtabular}
\end{table}
%%%%%%%%%%%%%%%%%%%%%%%%%%%%%%%%%%%%%%%%%%%%%%%%%%%%%%%%
%

The equilibrium properties of our model system were obtained by 
using standard canonical MC simulations following the Metropolis scheme. 
%\cite{Metropolis_JCP_1953,Allen_book_1987}
In detail, single particle (translational) moves were applied to all the counterions 
(i.e., anions and cations) and the monomers with an acceptance ratio of $50\%$.
%rod  
For the sake of efficiency of the spatial sampling, we also apply time to time 
``global'' translational displacements for the PA chain, 
where the {\it whole} chain is translated.
The same acceptance ratio of $50\%$ was applied there.
%It turns out that this chain move is crucial for the adsorption/desorption 
%feature.

All the simulation parameters are gathered in Table \ref{tab.simu-param}.
The classes of the simulated systems can be found in Table \ref{tab.simu-runs}.
For each class of systems characterized by a value of $\Xi$, we have systematically varied
the block length $M$ from 2 to 32 with intermediate values of $M=4,8,16$, leading
altogether to a set of 30 simulation runs.
The monomer concentration is set to 
$c_m=\frac{N_m}{L^2\tau} \approx 4.976 \times 10^{-4}a^{-3}$ 
leading to a PA volume fraction 
$\phi = \frac{4\pi}{3}c_m(a/2)^3 \approx 2.605 \times 10^{-4}$ 
(see also Table \ref{tab.simu-param}).
The total length of a simulation run is set to $3-4\times 10^6$ MC steps per particle.
Typically, the first $10^6$ MC steps were discarded, and 
about $2-3 \times 10^6$ MC steps were used to perform measurements. 

%\cite{note_barrier}.

%%%%%%%%%%%%%%%%%%%%%%%%%%%%%%%%%%%%%%%%%%%%%%%%%%%%%%%%
% TABLE 2
\begin{table}
\caption{
Classes of simulated systems characterized by their value of $\Xi=\frac{l_B}{\lambda}$. 
The number of counterions (cations and anions) ensuring
the overall electroneutrality of the system is not indicated.
The symbol $^{(\bigstar)}$ indicates {\it bulk} systems. 
}
\label{tab.simu-runs}
\begin{ruledtabular}
\begin{tabular}{lc}
 $\Xi$ & $\sigma_0L^2$ \\
 \hline
 $0^{(\bigstar)}$&
 $0$\\
 $0.9265$&
 $64$\\
 $1.853$&
 $128$\\
 $3.706$&
 $256$\\
 $7.412$&
 $512$
\end{tabular}
\end{ruledtabular}
\end{table}
%%%%%%%%%%%%%%%%%%%%%%%%%%%%%%%%%%%%%%%%%%%%%%%%%%%%%%%%

%%%%%%%%%%%%%%%%%%%%%%%%%%%%%%%%%%%%%%%%%%
\section{Results and discussion
 \label{Sec.Results}}
%%%%%%%%%%%%%%%%%%%%%%%%%%%%%%%%%%%%%%%%%%

%%%%%%%%%%%%%%%%%%%%%%%%%%%%%%%%%%%%%
\subsection{Preamble: Bulk behavior
 \label{Sec.chain_length}}
%%%%%%%%%%%%%%%%%%%%%%%%%%%%%%%%%%%%%

Before evoking the problem of a PA at a charged interface,
it is instructive to first consider the more simple bulk situation.
To our knowledge, the influence of the length ($M$) of the charged block   
on the PA chain-size has not been systematically investigated in the literature.
At all events, the understanding of the behavior of PA chains near a charged wall
necessitates a knowledge of the limiting bulk system.
To this end, we provide additional simulation data for the bulk situation (where $\Xi=0$).   
Thereby, the PA chain and their counterions are placed in a spherical cell of 
radius $R_{cell}=31.31a$ such that the monomer concentration $\frac{3N_m}{4\pi R_{cell}^3}$
remains identical to that studied in the slab geometry. The central $(N_m/2)^{\rm th}$ monomer along
the chain is held fixed and standard single particle moves as described above are applied.
The duration of the simulation is set to $10^8$ MCS leading to excellent statistics.   

\begin{figure}
\includegraphics[width = 8.0 cm]{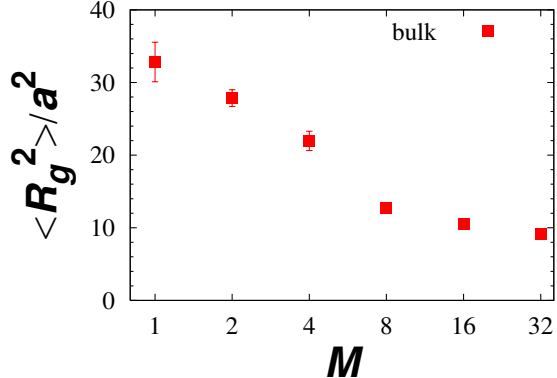}
\caption{
{\it Bulk} mean square radius of gyration $\langle {R_g^{(bulk)}}^2 \rangle$ 
as a function of the block length $M$.  
}
\label{fig.rg_bulk}
\end{figure}

To characterize the chain conformation the radius of gyration defined as
% 
%%%%%%%%%%%%%%%%%%%%
\begin{equation}
\label{eq.Rg2_bulk}
\left \langle {R_g^{(bulk)}}^2 \right \rangle =
\frac{1}{N_m}  \sum_{i=1}^{N_m} \left \langle \left( {\bf r}_i - {\bf r}_{cm}\right)^2 \right \rangle 
\end{equation}
%%%%%%%%%%%%%%%%%%%%
%
(where ${\bf r}_{cm}$ designates the center of mass of the chain) was monitored.
The results are presented in Fig. \ref{fig.rg_bulk}.
Simulations show unambiguously that the chain shrinks with increasing 
block length $M$. This behavior is the result of an increasing cohesion
between oppositely charged blocks with growing block charge. 
Similarly to what happens in an ionic crystal, upon increasing $M$
one concomitantly enlarges the coordination number (i.e., the average number of neighbors 
of opposite charge). 
%Consequently, bringing oppositely charged to contact reduces
%is energetically favorable.
For longer chains, it is expected that the chain size should saturate
from a certain threshold of $M$ due to excluded volume effects. 
Some support to this latter statement seems to be provided
by our simulations where the variation of the chain size is significantly slower
at large values of $M$ (see Fig. \ref{fig.rg_bulk}). 
Those preliminary findings will be useful to discuss our forthcoming results concerning
the PA chain behavior near a charged interface.

%%%%%%%%%%%%%%%%%%%%%%%%%%%%%%%%%%%%%%%%
% FIG 1
%%%%%%%%%%%%%%%%%%%%%%%%%%%%%%%%%%%%%%%%
%

%%%%%%%%%%%%%%%%%%%%%%%%%%%%%%%%%%%%%
\subsection{Behavior at weak  interface-ion Coulomb coupling
 \label{Sec.Qp64}}
%%%%%%%%%%%%%%%%%%%%%%%%%%%%%%%%%%%%%

In this part, we focus on the situation where the interface-ion Coulomb 
coupling is weak (i.e., essentially $\Xi=0.9265$).
Since the interface is (negatively) charged and the chain is {\it flexible}, 
different behaviors are expected for the positively and negatively charged blocks,
in contrast to what occurs in the bulk.
It is important to have in mind that the role of the flexibility of the chain 
is crucial here, since the stiff version of our PA would lead to a conformation
where the rod-like chain lies parallel to the normal $z$-direction of the interface
(with one end touching the interface) in a minimal energy configuration. 
%

%%%%%%%%%%%%%%%%%%%%%%%%%%%%%%%%%%%%%%%%
% FIG 2
\begin{figure}
\includegraphics[width = 8.0 cm]{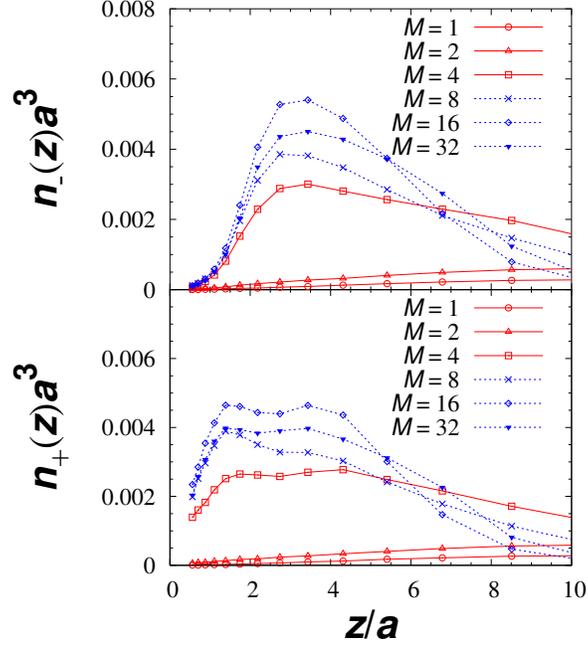}
\caption{Density profiles $n_{\pm}(z)^3$ at $\Xi=0.9265$ for various values of $M$.}
\label{fig.nz_Qp64}
\end{figure}
%%%%%%%%%%%%%%%%%%%%%%%%%%%%%%%%%%%%%%%%
%

%%%%%%%%%%%%%%%%%%%%%%%%%%%%%%%%%%%%%%%%
% FIG 3
\begin{figure}[b]
\includegraphics[width = 16.0 cm]{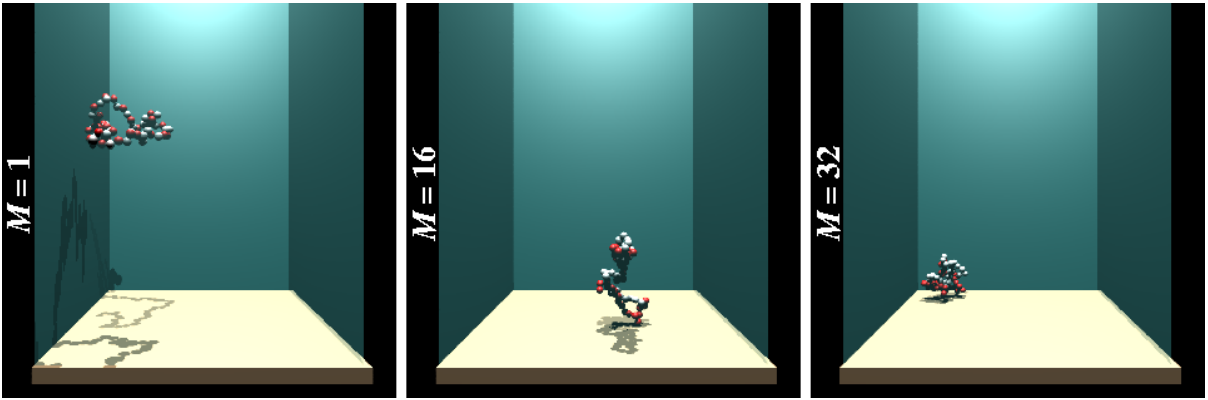}
\caption{Microstructure snapshots at $\Xi=0.9265$ for block lengths $M=1,16,32$
as indicated.
The red (white) monomers correspond to positive (negative) charges.
The little counterions are omitted for clarity.}
\label{fig.snap_Qp64}
\end{figure}
%%%%%%%%%%%%%%%%%%%%%%%%%%%%%%%%%%%%%%%%

To characterize the PA chain adsorption/depletion aspects, we have computed
the monomer density profiles $n_{\pm}(z)$ defined as 
% 
%%%%%%%%%%%%%%%%%%%%
\begin{equation}
\label{eq.nz}
n_{\pm}(z) = \sum_{i=1}^{N_m/2} \left \langle \delta \left(z-z_i^{(\pm)}\right) \right \rangle ,
\end{equation}
%%%%%%%%%%%%%%%%%%%%
%
and normalized  as follows
% 
%%%%%%%%%%%%%%%%%%%%
\begin{equation}
\label{eq.nz_norm}
\int ^{\tau-a/2}_{a/2} n_{\pm}(z) L^2 dz = \frac{N_m}{2},
\end{equation}
%%%%%%%%%%%%%%%%%%%%
%
where $+(-)$ applied to positively (negatively) charged monomers.
Our results are displayed in Fig. \ref{fig.nz_Qp64} and some 
microstructure snapshots are reported in Fig. \ref{fig.snap_Qp64}.
From a general viewpoint, the profiles of $n_{+}(z)$ and $n_{-}(z)$ in 
Fig. \ref{fig.nz_Qp64} are quite similar.
The reason for that observation is (i) the connectivity between oppositely
charged blocks and (ii) the weak Coulomb interface-block coupling.
Those density profiles $n_{\pm}(z)$ exhibit a peak from $M \geq 4$ which is a 
signature of PA adsorption (see Fig. \ref{fig.nz_Qp64}).
The disappearance of a marked peak in $n_+(z)$ can 
also be used as a criterion \cite{note_adsortion_criterion} for adsorption/depletion
transition in our simulation. 
Such a criterion was utilized by Shafir el al. in their theoretical 
study of adsorption and depletion of polyelectrolytes from oppositely charged
surfaces \cite{Shafir_JCP_2003}.
Near contact ($z/a \lesssim 1.5$) negatively charged monomers are strongly repelled from the wall
due to chain-entropy loss and wall-block electrostatic repulsion
[see $n_-(z)$ in Fig. \ref{fig.nz_Qp64}].
The positively charged monomers experience  a short range repulsion as well 
but this time {\it only} due to chain-entropy effects [see $n_+(z)$ in Fig. \ref{fig.nz_Qp64}].
This feature is well known for neutral chains \cite{DeGennes_Book_1979}, 
and was also reported by several authors for charged ones 
\cite{Varoqui_JPhysII_1993,Borukhov_Macromolecules_1998,Shafir_JCP_2003,Messina_PRE_2004}.  
The fact that the peak in $n_+(z)$ is considerably broadened (here at $M \geq 4$, see Fig. 
\ref{fig.nz_Qp64}) suggests that the adsorption is rather weak. 
This statement will be justified later where larger $\Xi$ coupling parameters are considered.
%\nu
It could be instructive to compare the contact value of $n_+(z \to a/2)$ to that
obtain in the case the where only the substrate's counterions are present 
\cite{note_nu}
[i.e., the well known and {\it exact} contact value from the Gouy-Chapmann theory 
$n_c^{(GC)}=(2\pi l_B \lambda^2)^{-1}=\frac{\Xi^2}{2 \pi l_B^3}$].
Introducing the dimensionless quantity $\nu$ defined as
% 
%%%%%%%%%%%%%%%%%%%%
\begin{equation}
\label{eq.nu}
\nu = \frac{n_+(z \to a/2)}{n_c^{(GC)}} 
= \frac{n_+(z \to a/2)}{\Xi^2 (2\pi l_B^3)^{-1}}, 
\end{equation}
%%%%%%%%%%%%%%%%%%%%
%
Fig. \ref{fig.nz_Qp64} shows that a typical value of $\nu$ in the adsorption regime is
roughly of the order of $\nu \approx \frac{0.002}{0.0288} \approx 0.07$.
This low value confirms that the adsorption at $\Xi=0.9265$ is (very) weak.
Close to the wall ($z/a \lesssim 4$), Fig. \ref{fig.nz_Qp64} indicates that $n_{\pm}$ 
increases systematically with $M$ if the special {\it diblock} case ($M=32$) is ignored.
The non-trivial behavior at $M=32$ is a result of  a subtle competition between 
the interface-block interaction and the block-block correlations. The latter favor {\it non-flat}
globular structures as well illustrated in Fig. \ref{fig.snap_Qp64} for $M=32$.
That effect can in turn lower the degree of adsorption and explain the non-trivial 
adsorption behavior reported in Fig. \ref{fig.nz_Qp64} at high $N_m$.

\begin{figure}
\includegraphics[width = 8.0 cm]{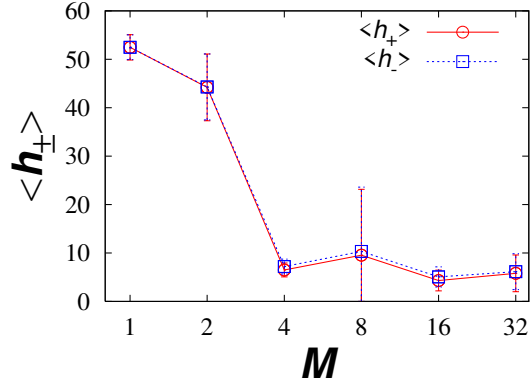}
\caption{
Reduced mean heights $\langle h_{\pm} \rangle$
as a function of block length $M$
}
\label{fig.h_Qp64}
\end{figure}

To further characterize the PA behavior at charged interfaces, 
we have measured the reduced mean height $\langle h_{\pm} \rangle$ defined
as

% 
%%%%%%%%%%%%%%%%%%%%
\begin{equation}
\label{eq.h}
\langle h_{\pm} \rangle = 
\int ^{\tau-a/2}_{a/2} n_{\pm}(z) \frac{z}{a} L^2 dz,
\end{equation}
%%%%%%%%%%%%%%%%%%%%
%
where $+(-)$ applied to positively (negatively) charged monomers.
Within this definition the mean height of the center of mass,
$\langle h_{cm} \rangle$, corresponds to  
$\langle h_{cm} \rangle = \frac{\langle h_{+} \rangle + \langle h_{-} \rangle}{2}$.
$\langle h_{\pm} \rangle$ as a function of $M$ (at $\Xi=0.9265$) are 
reported in Fig. \ref{fig.h_Qp64}.
One can distinguish two typical regimes (see Fig. \ref{fig.h_Qp64}):
(i) At short charged blocks ($M=1,2$) the values of $\langle h_{\pm} \rangle$
are such that $\langle h_{\pm} \rangle \approx \frac{\tau}{2a}$ 
(within the statistical uncertainties) so that the chain is fully {\it delocalized}
\cite{Messina_EPL_2006}.  (ii) For longer blocks ($M \geq 4$),
the mean heights decay by roughly one order of magnitude 
indicating an adsorption behavior that is consistent with 
the density profiles from Fig. \ref{fig.nz_Qp64}.    

%%%%%%%%%%%%%%%%%%%%%%%%%%%%%%%%%%%%%%%%
% FIG 4
%%%%%%%%%%%%%%%%%%%%%%%%%%%%%%%%%%%%%%%%
%

%%%%%%%%%%%%%%%%%%%%%%%%%%%%%%%%%%%%%%%%
% FIG 5
\begin{figure}[b]
\includegraphics[width = 8.0 cm]{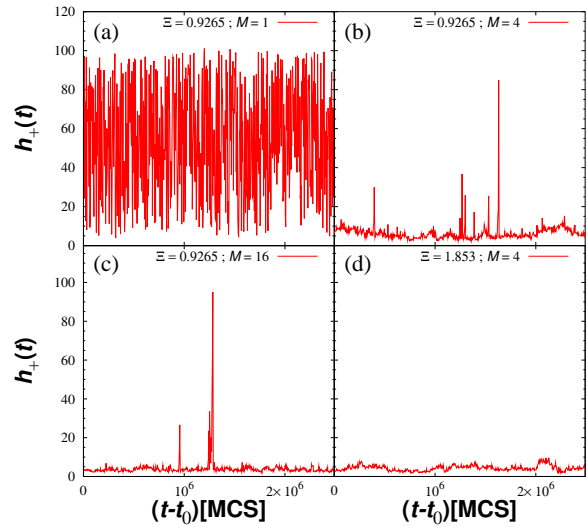}
\caption{
Reduced height $h_+(t)$ as a function of time for
different parameters as indicated. 
$t_0$ is a reference time.
}
\label{fig.h_time}
\end{figure}
%%%%%%%%%%%%%%%%%%%%%%%%%%%%%%%%%%%%%%%%
%

We now would like to address the problem of {\it desorption}.
The term ``desorption'' generally means that some adsorbed 
substance gets released, which implicitly involves a
dynamic process. For our present study, this latter feature can be 
best captured by monitoring the height of the  monomers as a function 
of time. For that issue, we define the reduced height 
of positive charges, $h_+(t)$, as follows
% 
%%%%%%%%%%%%%%%%%%%%
\begin{equation}
\label{eq.h_time}
h_{+} (t) = \frac{1}{a}  \frac{ \sum_{i}^{+} z_i(t),}{N_m/2},
\end{equation}
%%%%%%%%%%%%%%%%%%%%
%
where the symbol ``$+$'' in the sum means that only positively charged monomers are counted.
The results for  $h_+(t)$ for different set of parameters are shown in Fig. \ref{fig.h_time}. 
The panels (a), (b) and (c) from Fig. \ref{fig.h_time} correspond to the lowest value of
$\Xi=0.9265$ whereas panel (d) corresponds to a higher value of $\Xi=1.853$.
The case of weakest interface-block coupling [i.e., $\Xi=0.9265$ and $M=1$, see Fig. \ref{fig.h_time}(a)]
shows large fluctuations of amplitude $h_+^{(max)} - h_+^{(min)} \sim \tau/a$ that are a dynamic signature 
of the delocalization, in agreement with the value of the static quantity
$\langle h_{+} \rangle \approx \frac{\tau}{2a}$ found in Fig. \ref{fig.h_Qp64} for $M=1$.
Upon increasing the interface-block coupling (i.e., increasing $M$ and/or $\Xi$), 
Fig. \ref{fig.h_time} indicates that the probability of escaping from the adsorbed state
gets gradually weaker, as intuitively expected.
The fact that at sufficiently high interface-block coupling the desorption processes
become rare events, may in some cases considerably affect the quality of the statistical 
averaging as observed in Fig. \ref{fig.h_Qp64} for $M=8$.
Consequently, a perfect equilibration is not always guaranteed for some parameters.
Experimentalists are faced to the same problem when dealing with 
adsorption/desorption processes. 
This being said, we are confident that our simulations are reliable enough
to capture the main qualitative features of the behavior of non-random 
PA chains at charged surfaces.

%%%%%%%%%%%%%%%%%%%%%%%%%%%%%%%%%%%%%
\subsection{Behavior at intermediate and large  interface-ion Coulomb coupling
 \label{Sec.highQp}}
%%%%%%%%%%%%%%%%%%%%%%%%%%%%%%%%%%%%%

In this part, larger interface-ion Coulomb couplings are considered corresponding
to $\Xi=1.853,3.706$ and $7.412$ (see Table \ref{tab.simu-runs}) 
with varying block lengths $M$.
Many observables are going to be analyzed in detail and those can be classified
into two main categories: (i) monomer concentration profiles 
and related quantities and (ii) chain conformation properties.

%%%%%%%%%%%%%%%%%%%%%%%%%%%%%%%%%%%%%
\subsubsection{Monomer distribution
 \label{Sec.legendre_Qp-64}}
%%%%%%%%%%%%%%%%%%%%%%%%%%%%%%%%%%%%%

%%%%%%%%%%%%%%%%%%%%%%%%%%%%%%%%%%%%%%%%
% FIG 6
\begin{figure}
\includegraphics[width = 15.0 cm]{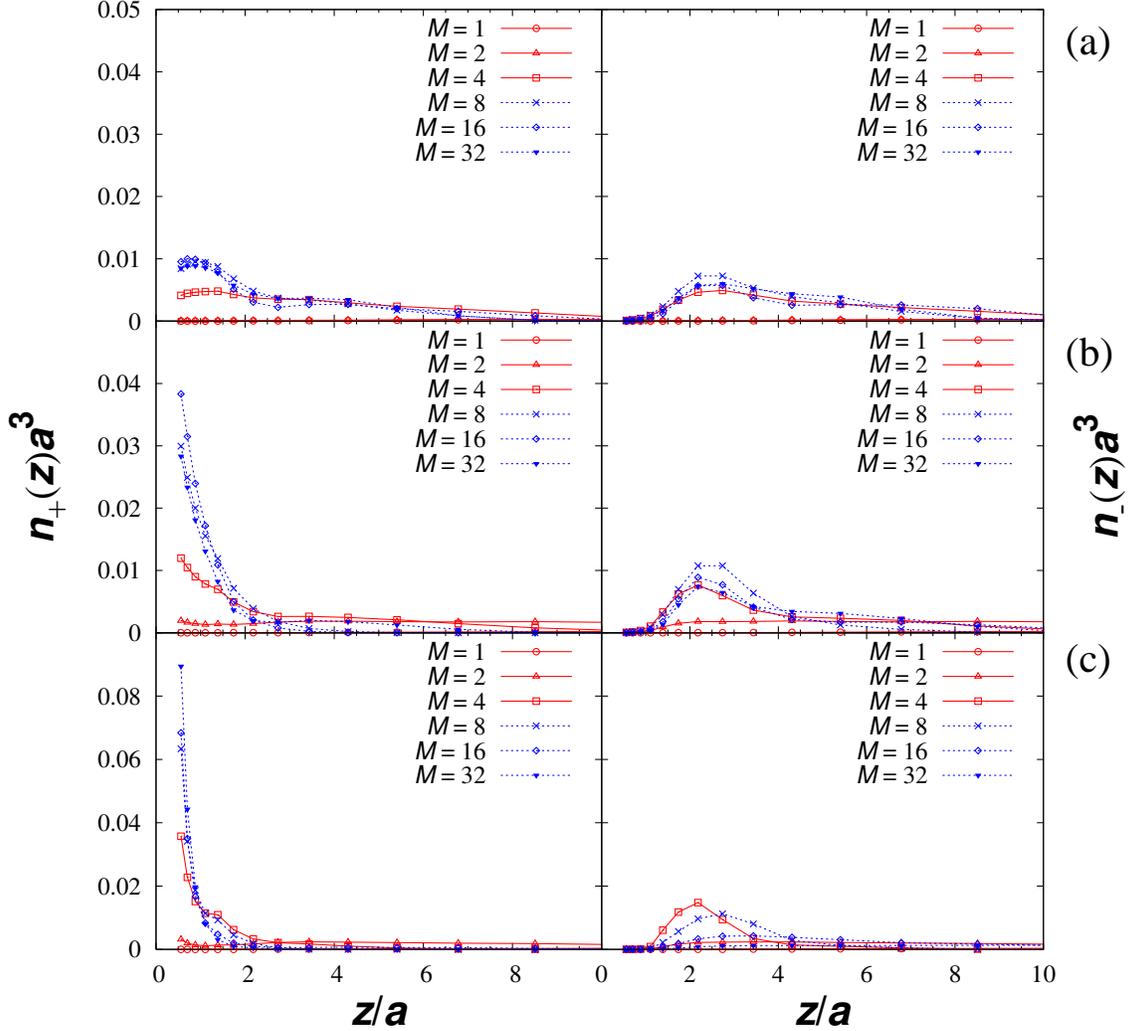}
\caption{Density profiles $n_{\pm}(z)^3$ 
for different block lengths $M$. The values of $\Xi$ increase from top to bottom:
(a) $\Xi=1.853$ 
(b) $\Xi=3.706$
(c) $\Xi=7.412$.
}
\label{fig.nz_HighQp}
\end{figure}
%%%%%%%%%%%%%%%%%%%%%%%%%%%%%%%%%%%%%%%%
%

%%%%%%%%%%%%%%%%%%%%%%%%%%%%%%%%%%%%%%%%
% FIG 7
\begin{figure}
\includegraphics[width = 14.0 cm]{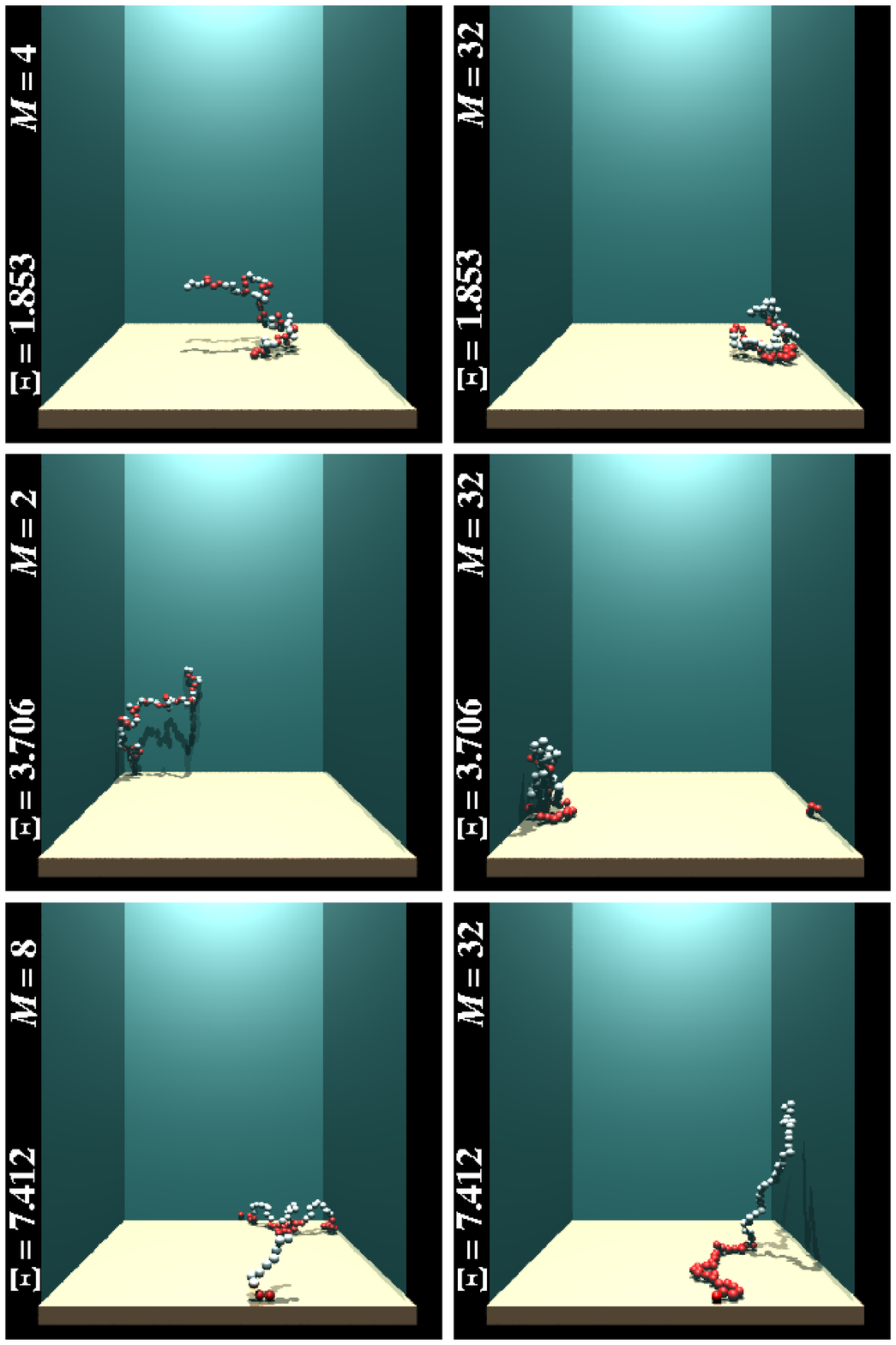}
\caption{Microstructure snapshots at different values of $\Xi$ and 
$M$ as indicated. Note that $\Xi$ increases from top to bottom.
The red (white) monomers correspond to positive (negative) charges.
The little counterions are omitted for clarity.}
\label{fig.snap_HighQp}
\end{figure}
%%%%%%%%%%%%%%%%%%%%%%%%%%%%%%%%%%%%%%%%
%

The monomer density profiles are depicted in Fig. \ref{fig.nz_HighQp}.
and some corresponding microstructure snapshots are sketched 
in Fig. \ref{fig.snap_HighQp}. 
Let us first discuss the behavior of the positively charged monomers
which is relevant for the adsorption, since without those no
adsorption at all would occur.
At $\Xi=1.853$, the density profiles of the positively charged
monomers exhibit similarity to that at $\Xi=0.9265$ 
(compare Fig. \ref{fig.nz_HighQp} and Fig. \ref{fig.nz_Qp64}),
where especially the repulsive chain-entropy effects are present.
An adsorption behavior is detected here for $M \geq 4$ 
[see Fig. \ref{fig.nz_HighQp}(a)].
This onset of adsorption can also be nicely visualized 
by comparing the snapshot of Fig. \ref{fig.snap_HighQp}
(at $M=4$ and $\Xi=1.853$) to that of Fig. \ref{fig.snap_Qp64}
(at $M=1$ and $\Xi=0.9265$) corresponding to the non-adsorbing case. 
However, at given $M$, more positively charged monomers can be found near the 
oppositely charged wall at $\Xi=1.853$ than at $\Xi=0.9265$ 
[roughly by a factor 2, compare Fig. \ref{fig.nz_HighQp}(a) and Fig. \ref{fig.nz_Qp64}].
At higher wall-ion couplings (i.e., $\Xi=3.706$ and $\Xi=7.412$), the scenario
gets qualitatively different (see Fig. \ref{fig.nz_HighQp}), where the
$n_+(z)$-profiles are now {\it monotonic} in contrast to what was found
at $\Xi=0.9265$ (see Fig. \ref{fig.nz_Qp64}) and $\Xi=1.853$. 
Physically, this means that the electrostatic effects overcome
the chain-entropy ones at sufficiently large $\Xi$.
This feature can also be seen as signature of {\it strong} adsorption.
A closer look at Fig. \ref{fig.nz_HighQp} reveals another
interesting behavior, namely at our largest value of $\Xi=7.412$,
the density $n_+(z)$ near contact {\it increases systematically} with
growing block length $M$, in contrast to what happens at
lower $\Xi$ (see also Fig. \ref{fig.nz_Qp64}).
This phenomenon can be explained by the strong wall-block
electrostatic interaction that overcomes the cohesive 
block-block correlations. Those both effects grow with
$M$ in a non trivial way.

As far as the local concentration of negatively charged monomers $n_-(z)$
is concerned, Fig. \ref{fig.nz_HighQp} indicates a complicated behavior
as a function of $\Xi$ and $M$ in the adsorption regime (here $M \geq 4$).
More explicitly, at $\Xi=1.853$ the maxima reached by $n_-(z)$ at
long enough block length (here $M \geq 8$) are smaller than those of
$n_+(z)$ [see Fig. \ref{fig.nz_HighQp}(a)], 
in contrast to what was reported at $\Xi=0.9265$ (see Fig. \ref{fig.nz_Qp64}).
This behavior is due to the stronger wall-block correlation existing
at $\Xi=1.853$. It is to say that the positively charged blocks
get more attracted towards the negatively charged wall at higher $\Xi$ 
and concomitantly the negatively ones get more repelled from it.
The density profile $n_-(z)$ at $\Xi=3.706$ is qualitatively similar
to that obtained at $\Xi=1.853$ (see Fig. \ref{fig.nz_HighQp}),
and a truly different behavior is found at $\Xi=7.412$.
There, in the strong adsorption regime 
(i.e., $\Xi=7.412$ with $M \geq 4$), Fig. \ref{fig.nz_HighQp}(c)
interestingly indicates that only chains with intermediate block 
length $M=4,8$ can exhibit (significant) maxima in $n_-(z)$.
The physical interpretation of this feature is that at high
$M$ the repulsion between the wall and the negatively charged blocks
is so strong that, despite of the chain connectivity, 
the negatively charged get strongly repelled from the wall leading to 
formation of {\it loops} and {\it tails}.
This mechanism explains also why $n_-(z)$, at $\Xi=7.412$, is higher
for $M=4$ than for $M=8$ in Fig. \ref{fig.nz_HighQp}(c).
All those relevant findings at $\Xi=7.412$ are convincingly illustrated 
in Fig. \ref{fig.snap_HighQp} (at  $\Xi=7.412$) where 
a ``{\it caterpillar}'' structure is developed
for $M=8$,  whereas a ``L'' structure rises for $M=32$.
Note that the former structure was also theoretically reported by Dobrynin et al.
\cite{Dobrynin_Macromol_1997} for symmetric PAs that 
was referred to as the ``pancake regime''.
On the other hand, the ``L'' structure deserves some further comments.
Naively one could think that solely the strong electrostatic 
interface-block interactions are sufficient to explain this structure.
Basic electrostatics \cite{note_L_structure} show that this idea is wrong, 
and one must take into account the energy gain upon approaching the two oppositely charged blocks 
when passing from a stretched configuration (perpendicular to the interface) to the ``L'' structure.
Hence, the ``L'' structure is again the result of interface-block 
interactions and block-block correlations.

%%%%%%%%%%%%%%%%%%%%%%%%%%%%%%%%%%%%%%%%
% FIG 8
\begin{figure}
\includegraphics[width = 8.0 cm]{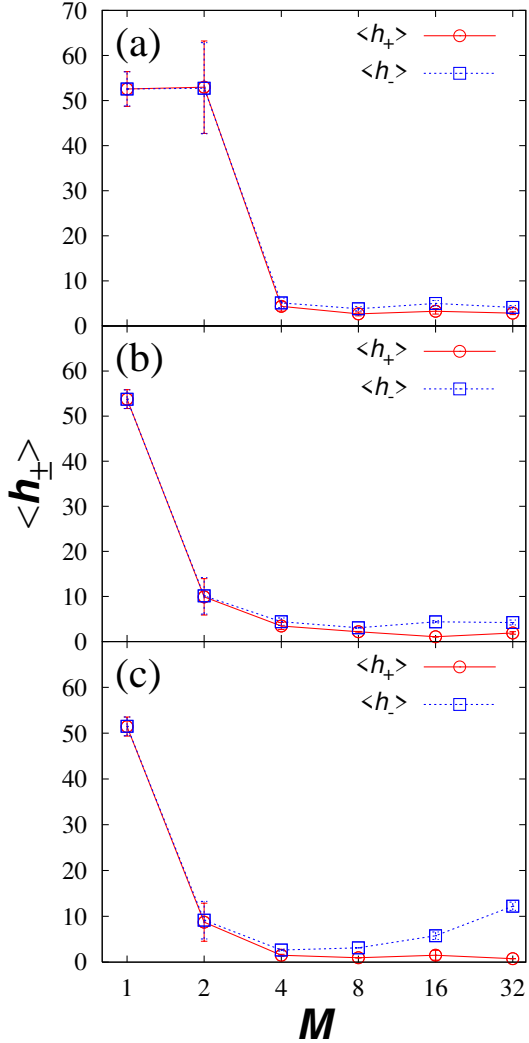}
\caption{
Reduced mean heights $\langle h_{\pm} \rangle$
as a function of block length $M$ for various values of $\Xi$ increasing from top to bottom:
(a) $\Xi=1.853$ 
(b) $\Xi=3.706$
(c) $\Xi=7.412$.
}
\label{fig.h_HighQp}
\end{figure}
%%%%%%%%%%%%%%%%%%%%%%%%%%%%%%%%%%%%%%%%
%

To gain further insight into the monomer distribution, we
have plotted the mean heights $\langle h_{\pm} \rangle$ as
a function of $M$ in Fig. \ref{fig.h_HighQp}.
For all values of $\Xi$, a {\it delocalized} state is reported
at $M=1$ as signaled by the value of 
$\langle h_{\pm} \rangle \approx \tau/2$ (see Fig. \ref{fig.h_HighQp}).
That non-adsorption at $M=1$ is due to the very weak polarizability of the chain there.
The problem of chain polarization [see Eq. \eqref{eq.polarization}] 
through the electric field stemming from the interface will be discussed in more detail later.
Given that our highest value $\Xi=7.412$ would correspond to 
$|\sigma_0 e| \approx 1e/(43 {\rm \AA}^2)$ 
(see also Table \ref{tab.simu-param} for the chosen values of system parameters),
which represents already the upper limit of experimentally accessible surface charge densities,
we expect that no adsorption should occur in reasonable experimental conditions when $M=1$. 
At $M=2$, only the case $\Xi=1.853$ remains clearly delocalized 
(see Fig. \ref{fig.h_HighQp}).
To rigorously assess the problem of localization/delocalization for 
the other remaining systems (especially for $M=2$ at higher $\Xi$), 
a more appropriate investigation involving  a systematic variation of the box height 
$\tau$ is required. Thereby if $\langle h_{\pm} \rangle \to \infty$ while $\tau \to \infty$,
then the chain is said delocalized, otherwise it is localized.
Noticing that $\langle h_{+} \rangle - \langle h_{-} \rangle$ represents
the (reduced) {\it dipolar moment} $\langle p \rangle$ of the chain given by
% 
%%%%%%%%%%%%%%%%%%%%
\begin{equation}
\label{eq.polarization}
\langle p \rangle = \left \langle \sum_{i=1}^{N_m}  Z_i z_i \right \rangle 
                  = N_m \frac{\langle h_{+} \rangle - \langle h_{-} \rangle}{2}, 
\end{equation}
%%%%%%%%%%%%%%%%%%%%
%
a closer look at 
Fig. \ref{fig.h_HighQp} reveals (at given $M \geq 4$) that the adsorbed chain 
gets more and more polarized with increasing $\Xi$.
The chain polarization is particularly vivid at high wall-block 
couplings where it increases with $M$ [see Fig. \ref{fig.h_HighQp}(c)].
Concomitantly, this $M$-enhanced polarization effect yields a strong {\it non-monotonic} 
behavior for $\langle h_{-} \rangle$ at $\Xi=7.412$ [see Fig. \ref{fig.h_HighQp}(c)].
Nonetheless, we stress the point that, in general for intermediate and weak ion-wall couplings $\Xi$, 
the polarization behavior as a function of $M$
is non-trivial since $\langle p \rangle$ is also sensitive to the block-block correlations.

%%%%%%%%%%%%%%%%%%%%%%%%%%%%%%%%%%%%%
\subsubsection{Conformational properties
 \label{Sec.conformation}}
%%%%%%%%%%%%%%%%%%%%%%%%%%%%%%%%%%%%%

%%%%%%%%%%%%%%%%%%%%%%%%%%%%%%%%%%%%%%%%
% FIG 9
\begin{figure}
\includegraphics[width = 8.0 cm]{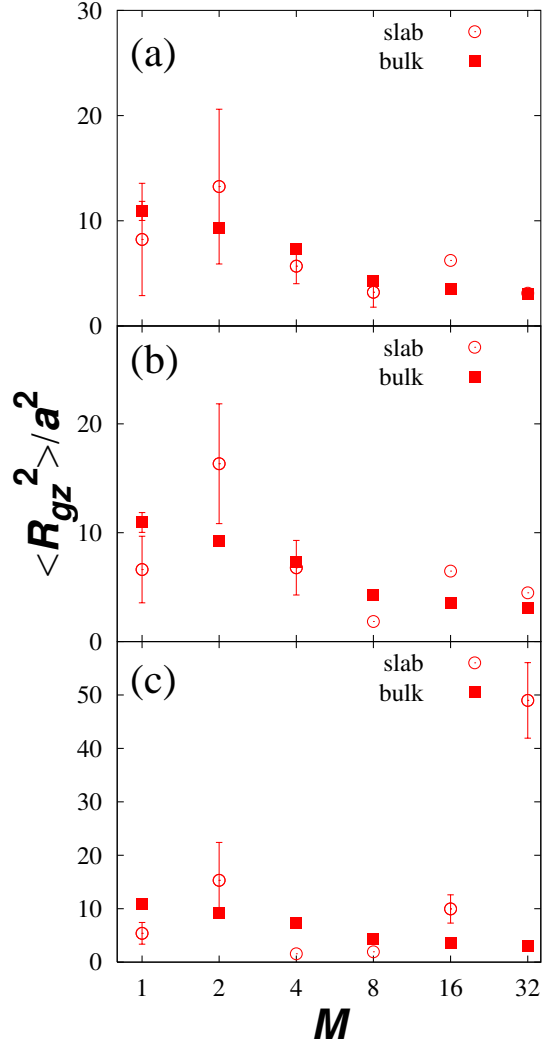}
\caption{
Mean square radii of gyration $\langle R_{gz}^2 \rangle$ (slab) 
and  $\langle {R_{gz}^{(bulk)}}^2 \rangle$ (bulk) 
as a function of block length $M$ for various values of $\Xi$ increasing from top to bottom:
(a) $\Xi=1.853$ 
(b) $\Xi=3.706$
(c) $\Xi=7.412$.  
}
\label{fig.rg_slab}
\end{figure}
%%%%%%%%%%%%%%%%%%%%%%%%%%%%%%%%%%%%%%%%
%

A suitable quantity assessing the conformation of the chain is
provided by the $z$-component $\langle R_{gz}^2 \rangle$  of the 
squared radius of gyration of the chain that is defined as follows
% 
%%%%%%%%%%%%%%%%%%%%
\begin{equation}
\label{eq.Rg2_z}
\langle R_{gz}^2 \rangle =
\frac{1}{N_m}  \sum_{i=1}^{N_m} \left \langle \left( z_i - z_{cm}\right)^2 \right \rangle, 
\end{equation}
%%%%%%%%%%%%%%%%%%%%
%
where $z_{cm}$ designates the $z$-component of the center of mass of the chain.
The square root $\sqrt{\langle R_{gz}^2 \rangle}$  can also be seen as the 
{\it thickness} of the chain in the adsorbed state.
It is instructive to compare those values to those obtained in the bulk case 
(see Fig. \ref{fig.rg_bulk}). To this end, we introduce the following dimensionless quantity
% 
%%%%%%%%%%%%%%%%%%%%
\begin{equation}
\label{eq.eta}
\eta = \frac{\langle R_{gz}^2 \rangle}{\langle {R_{gz}^{(bulk)}}^2 \rangle}, 
\end{equation}
%%%%%%%%%%%%%%%%%%%%
%
where $\langle {R_{gz}^{(bulk)}}^2 \rangle = \frac{1}{3}\langle {R_g^{(bulk)}}^2 \rangle$.
Thereby, $\eta < 1$ signals an interface-induced compression whereas
$\eta > 1$ indicates an interface-induced stretching.
Results for $\langle R_{gz}^2 \rangle$ are presented in Fig. \ref{fig.rg_slab}.
Despite of the rather large error bars appearing for the data points  
corresponding to the depletion/weak adsorption regimes, our simulations show that
there is a systematic compression/stretching transition when passing from $M=1$ to $M=2$. 
This non-trivial phenomenon can be rationalized by considering the {\it stiff} limiting case.
Indeed in the situation of a stiff rod-like PA, the minimal energy configuration
(ignoring the counterions) would correspond to a chain ``standing'' on the wall
with the positively charged end touching the interface, as previously discussed.
Thereby, it is straightforward to show that the energy gain increases with $M$
(i.e., the dipolar moment of the chain).
Hence at finite temperature, the rod-like will adopt an average orientation that
gradually tends to be parallel to the $z$-direction when $M$ is enlarged.
Now, this idea can be qualitatively applied to flexible chains as long as the
block-block correlations are not too strong, which is qualitatively the case for $M=1,2$ 
(see Fig. \ref{fig.rg_bulk}).
Our scenario is perfectly illustrated by the snapshot in Fig. \ref{fig.snap_HighQp}
at $\Xi=3.706$ and $M=2$.
This stretching at $M=2$ implies also larger values of $\langle R_{gz}^2 \rangle$ 
than at $M=1$ (see Fig. \ref{fig.rg_slab}). 
Note that this structure was also theoretically reported by Dobrynin et al.
\cite{Dobrynin_Macromol_1997} for symmetric PAs that 
was referred to as the ``pole regime''.

For longer blocks the situation is more complicated since there is a subtle 
competition between block-block correlations and block-wall ones. 
At intermediate block lengths ($4 \leq M  \leq 8$) the chain
is (again) in the compression regime (see. Fig. \ref{fig.rg_slab}), whereas
by further increasing $M$ the chain gets stretched.
A possible exception to this scenario is the diblock case ($M=32$) at $\Xi=1.853$  
where $\eta \approx 1$ due to the strong block-block correlations and the rather 
weak interface-ion coupling there (see. Fig. \ref{fig.rg_slab}).
The ``L'' structure adopted by a diblock polymer chain at $\Xi=7.412$ 
(see Fig. \ref{fig.snap_HighQp}) leads to a strong stretching stemming 
from the negatively charged block (see. Fig. \ref{fig.rg_slab}).

\begin{figure}
\includegraphics[width = 15.0 cm]{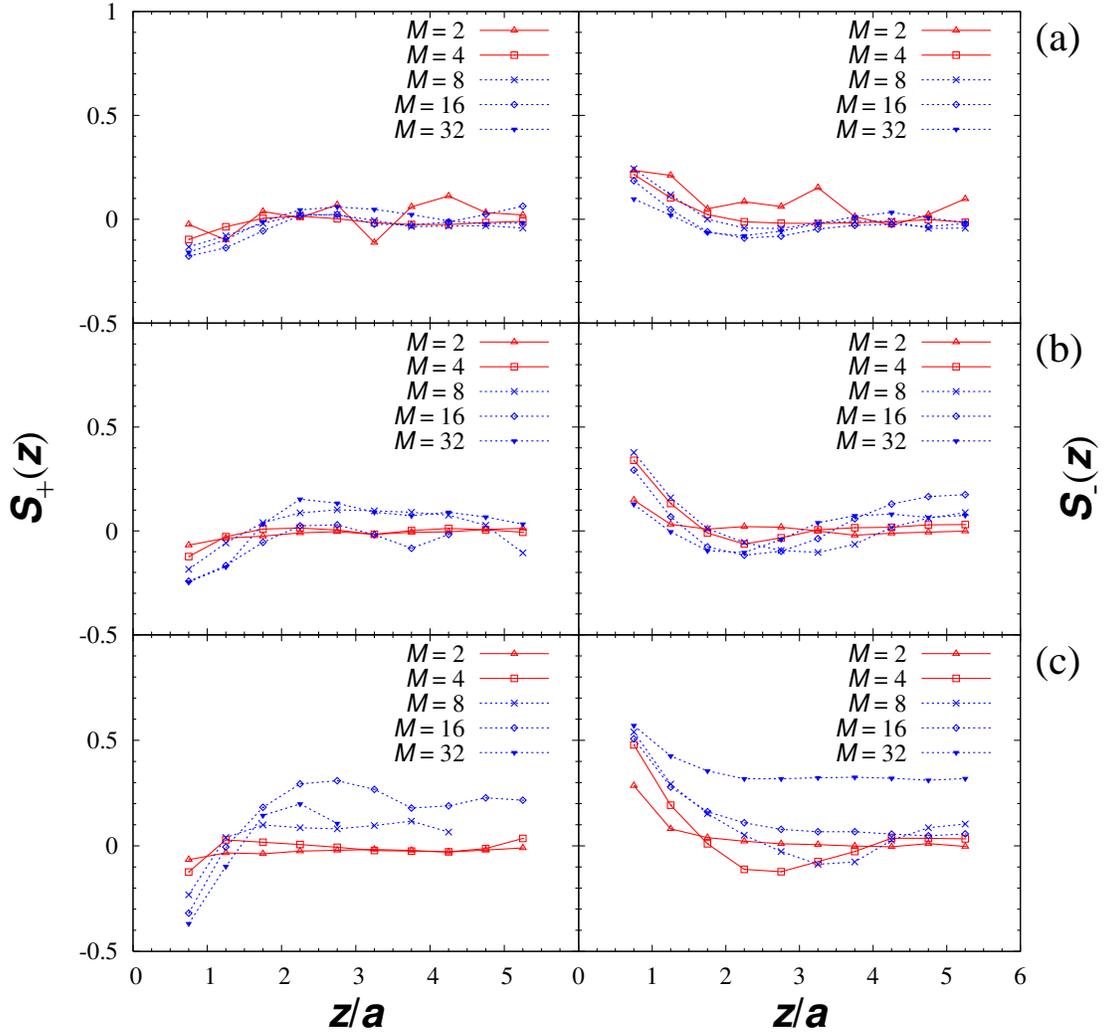}
\caption{
Average bond-wall orientation correlation function $S_{\pm}(z)$
for different block lengths $M$. The values of $\Xi$ increase from top to bottom:
(a) $\Xi=1.853$ 
(b) $\Xi=3.706$
(c) $\Xi=7.412$.  
}
\label{fig.Sz}
\end{figure}

The chain conformation can be further characterized by evaluating
the bond-wall orientation correlation. 
To quantify this order parameter we monitor the following second order Legendre
polynomial
%
%%%%%%%%%%%%%%%%%%%%
\begin{equation}
\label{eq.S_z}
S_{\pm}(z)  = 
\frac{
%num 
\left \langle
\displaystyle 
\sum_{i=1}^{N_{bond}} 
\frac{1}{2}  \left[ 
3\left( \frac{{\bf b}_{i}^{(\pm)} \cdot {\bf e}_{z}}{\left | {\bf b}_{i}^{(\pm)} \right |}  \right)^2
-1 
\right]
\delta \left( z - \tilde z_{i}^{(\pm)}\right)
\right \rangle 
}
%den
{
\left \langle
\displaystyle \sum_{i=1}^{N_{bond}} 
\delta \left( z - \tilde z_{i}^{(\pm)} \right)
\right \rangle
} 
\end{equation}
%%%%%%%%%%%%%%%%%%%%
%
where ${\bf b}_{i}^{(\pm)}={\bf r}_{i+1}^{(\pm)}-{\bf r}_{i}^{(\pm)}$ 
(with ${\bf r}_{i}^{(\pm)}=x_i^{(\pm)}{\bf e}_{x} + y_i^{(\pm)}{\bf e}_{y} + z_i^{(\pm)}{\bf e}_{z}$ )
stands for the bond vector linking only positively $(+)$ charged monomers or only 
negatively $(-)$ charged monomers, $N_{bond}=\frac{N_m(M-1)}{2M}$ is the total number 
of ``positive'' (or ``negative'') bonds, and
$\tilde z_{i}^{(\pm)} = \min \left(z_{i}^{(\pm)}, z_{i+1}^{(\pm)} \right)$.
Hence, $S_{\pm}(z)$ defined for $M \geq 2$ reaches the values $-\frac{1}{2}$, 0, and $+1$ for bonds 
that are perpendicular to the $z$-axis, randomly oriented, and parallel to the 
$z$-axis, respectively.
Results for $S_{\pm}(z)$ are sketched in Fig. \ref{fig.Sz}.
Let us first analyze $S_{+}(z)$ where bonds link positively charged monomers.
At contact ($z \lesssim a$), it can be seen from Fig. \ref{fig.Sz} that
the positive bonds typically tend to lie the more parallel to the wall the larger the block lengths, 
as signaled by negatively stronger values of $S_{+}(z)$. 
Moreover, the wall-bond degree of parallelism is enhanced when $\Xi$ is increased
at contact (see Fig. \ref{fig.Sz}).
Sufficiently far away from the wall, bonds get typically randomly oriented
as signaled by a near 0 value reached by  $S_{+}(z)$ (see Fig. \ref{fig.Sz}).
Nonetheless, in the regime of strong adsorption 
(i.e., $\Xi=7.412$ and for $M \geq 4$), our simulation data show some deviation from
that rule [see Fig. \ref{fig.Sz}(c)]. This is merely due to the fact that at strong adsorption,
positively charged monomers lying ``far away'' from the wall constitute rare events and 
therefore our results are not reliable for such high wall separations.

%%%%%%%%%%%%%%%%%%%%%%%%%%%%%%%%%%%%%%%%
% FIG 10
%%%%%%%%%%%%%%%%%%%%%%%%%%%%%%%%%%%%%%%%
%

Concerning the bonds linking negatively charged monomers, an inspection of Fig. \ref{fig.Sz}
reveals that the $S_{-}(z)$-profiles can be seen qualitatively as 
the mirror reflexion of the $S_{+}(z)$-profiles about the axis $S_{\pm}(z)=0$. 
Still on a qualitative level, this mirror symmetry reflects simply the 
fact that positive bonds tend to get parallel to the wall whereas negative ones 
tend to stand perpendicular to the wall. 
This scenario is especially vivid at high $\Xi=7.412$ as can be seen in Fig. \ref{fig.Sz}(c) 
and in the snapshots from Fig. \ref{fig.snap_HighQp} taken at $\Xi=7.412$. 
In particular, for the diblock case ($M=32$), the snapshot  [see Fig. \ref{fig.snap_HighQp}(c)] 
reveals that the tail is stretched leading to a  strong  and {\it long-ranged} wall-bond 
orientation correlation with a typical correlation length that should be proportional to $Ma$.

%%%%%%%%%%%%%%%%%%%%%%%%%%%%%%%%%%%%%
\section{Concluding remarks
 \label{Sec.summary}}
%%%%%%%%%%%%%%%%%%%%%%%%%%%%%%%%%%%%%

We have presented a systematic set of MC simulations to address the behavior of
a single flexible charge-ordered polyampholyte chain near a charged planar interface.
The chain length was fixed and the constitutive blocks were
highly charged with two consecutive blocks being oppositely charged. 
The influence of block length $M$ and interface-ion Coulomb coupling
$\Xi$ on the overall chain conformation and adsorption properties were analyzed.
Several relevant quantities were studied such as the local monomer concentration,
the mean height of the centers of the positive and negative charges, 
the perpendicular component of the mean squared radius of gyration, 
the interface-bond orientation correlation function, 
and illustrative microstructural snapshots.  
Based on those results, our main findings can be summarized as follows:
%%%%%%
\begin{itemize}
% M=1
\item {\it No adsorption} occurs at monomer-sized blocks due to the weak chain polarization there.
% adsorption
\item Chain adsorption can be obtained by increasing the {\it interface-block} coupling,
      i. e., increasing $M$ {\it and/or} $\Xi$.
% compression-stretching
\item By taking the bulk case as reference, compression-stretching transition 
      sets in when passing from monomer-sized blocks to dimer-sized ones.
      The same behavior applies to the adsorption regime when passing 
      from intermediate values of $M$ to high values of $M$. 
\item Formations of tails and loops manifest only for strong
      interface-block couplings.
\end{itemize}
%%%%%%

These findings can be possibly useful to provide some hints to recent experiments
\cite{Mahltig_JColIntSc_2001} where the (lateral) hydrodynamic diameter of 
adsorbed diblock polyampholitic structures on silicon substrates 
and that in the solution were measured by light scattering techniques. 
Consequently, the qualitative behavior of $\eta$ [Eq. \eqref{eq.eta}] 
is experimentally specified (see Fig. 4 in Ref. \cite{Mahltig_JColIntSc_2001}). 
Upon varying the pH Mahltig et al. \cite{Mahltig_JColIntSc_2001} were able to gradually
tune the net charge (from negative to positive passing through the isoelectric point) 
of the polyampholyte and that of the interface, 
and observed a similar compression-stretching scenario to that of our simulations.

%%%%%%%%%%%%%%%%%%%%%%%%%%%%%%%%%%%%%
%Acknowledgments
%%%%%%%%%%%%%%%%%%%%%%%%%%%%%%%%%%%%%
\acknowledgments 
The author thanks H. L\"owen for useful discussions.
Financial support from DFG via LO418/12 and SFB TR6 is also acknowledged.

%\bibliographystyle{epj}
%\bibliography{pe}

\begin{thebibliography}{24}
\bibitem{note_HS}
Only the monomer-monomer excluded volume interaction was not modelled by a
  hard-sphere potential. There, a purely repulsive Lennard-Jones potential was
  used.
\bibitem{note_adsortion_criterion}
Note that in our case where we have only one chain and no contact with a bulk
  reservoir, so that the monomer density profile will necessarily tend to zero.
  This means that the density will always show a peak (although very small in
  practice in case of depletion).
\bibitem{note_nu}
Note that for $\Xi=0.9265$ the number of charged monomers is identical to that
  of the substrate's counterions making the comparison meaningful.
\end{thebibliography}

\end{document}